\documentclass[doublecol]{epl2} 
\usepackage{amsmath,amsbsy,graphicx}
\title{Dynamics of an asymmetric bilayer lipid membrane in a viscous solvent}
\shorttitle{Dynamics of an asymmetric bilayer lipid membrane} 

\author{R.J. Bingham\inst{1} \and S.W. Smye\inst{2} \and P.D. Olmsted\inst{3}}
\shortauthor{R.J. Bingham \etal}

\institute{                    
  \inst{1} York Centre for Complex Systems Analysis, University of York, York, YO10 5GE, UK\\
  \inst{2} Academic Division of Medical Physics, University of Leeds, Leeds, LS2 9JT, UK \\
  \inst{3} Institute for Soft Matter Synthesis \& Metrology, Department of Physics, Georgetown University, Washington DC, USA
}
\abstract{Bilayer lipid membranes (BLMs) are an essential component of many biological systems, forming a functional barrier between the cell and the surrounding environment. When the membrane relaxes from a structural perturbation, the dynamics of the relaxation depends on the bilayer structure.  We present a model of a BLM in a viscous solvent, including an explicit description of a `thick' membrane, where the fluctuations in the thickness of a monolayer leaflet are coupled to changes in the lipid density within that monolayer.  We find dispersion relations describing three intuitive forms of bilayer motion, including a mode describing motion of the intermonolayer surface not noted previously in the literature.  Two intrinsic length scales emerge that help characterise the dynamics; the well known Saffman-Delbr\"{u}ck length and another, $\ell_{r}$, resulting from the intermonolayer friction.  The framework also allows for asymmetry in the BLM parameters between the monolayer leaflets, which is found to couple dynamic modes of bilayer motion.}

\begin{document}

\maketitle

\section{Introduction}
Bilayer lipid membranes (BLMs) are ubiquitous in nature and common in practical applications.  The complexity of biological membranes is typified by the cell membrane, a complex dynamic mix of proteins and lipids that act as the gatekeeper to the cell, the lipids additionally providing an active host for membrane proteins \cite{engelmanN2005}.  The double layer structure of a BLM allows for complex equilibrium conformations and rich dynamical behaviour \cite{lipowskynature1991}.  The BLM's role as a boundary between two diverse environments means that the membrane composition will rarely be symmetric across the monolayer leaflets that comprise the bilayer; the two membrane surfaces may have different functional requirements.  An asymmetry in composition will lead to an asymmetry in behaviour, which can be used as a control mechanism in biological processes \cite{Nelsonbook2004,fournierPRL2009,sensPRL2004}.  The symmetry of the membrane can also be broken by the imposition of an external field, such as the electric fields used in electroporation \cite{binghamPRE2010}.  The ways in which asymmetry may effect and even assist the biological functions of a BLM is as yet unclear.  Equally as unclear is the complex dynamic behaviour that can emerge in an asymmetric BLM. 

At the simplest level, a BLM can be modelled mathematically as a thin sheet embedded in fluid \cite{KramerJCP1971,caiPRE1995}, as employed by Brochard \& Lennon to model the fluctuations of red blood cell membranes \cite{BrochardJDP1975}.  A thin sheet can only generate a single hydrodynamic mode, where the displacement of the membrane is resisted by the curvature and tension of the membrane and energy is dissipated in the surrounding fluid.   A similar approach of coupled thin sheets has been used by Lu and Cates to model the hydrodynamics of surfactant films \cite{luL1995}.

Seifert \& Langer \cite{SeifertEPL1993,SeifertBC1994} expanded the description to a BLM comprising two compressible monolayer leaflets coupled to the bending of the bilayer.  The movement of lipids within a monolayer is restricted by intermonolayer friction, the viscosity of the surrounding solvent and the viscosity within a monolayer.  Seifert \& Langer's model considers three modes of motion; a height mode (as previously modelled by Brochard \& Lennon \cite{BrochardJDP1975}), an average density mode, where the monolayer densities move in phase, and a density difference mode, where the monolayer densities move out of phase.  The average density mode is not considered within the dissipative dynamic framework and is considered a purely propagating mode, which leaves combined dispersion relations for the remaining dynamic modes, where the dominant relaxation mode depends on the length scale of the bilayer movement.  This model shows quantitative agreement with experimental work \cite{pfeifferEPL1993,arriagaEPJE2010,hirnFD1999,monroyPRL2009} and has subsequently become the most widely used framework for studies of BLM dynamics.  H\"{o}mberg \& M\"{u}ller \cite{hombergEPL2012} generalised the Seifert \& Langer theory to include the bilayer's inertia and surface tension although the inclusion of inertial effects does not improve the agreement with experimental results since most nanoscale BLM undulations occur at vanishing Reynolds number.  Coarse-grained simulation models of membranes allows for the direct calculation of fluctuation spectra, where the Seifert \& Langer framework has again been used to effectively describe the fluctuations observed \cite{OtterBioJ2007,shkulipaPRL2006}.

Recent computational \cite{lindahlBioJ2000,watsonJCP2011} and experimental \cite{woodkaPRL2012,nagaoSM2011,PottEPL2002} work has highlighted the role fluctuations in the membrane thickness can play in the dynamics.  Thickness fluctuations are not explicitly considered in the Seifert \& Langer model, as the BLM is physically represented as a thin sheet.  Fluctuations in the lipid density in each leaflet are analogous to thickness fluctuations, but neglect the movement of fluid due to thickness changes, which will affect the dynamics.  Advances in computer simulation have also led to debate around the Seifert \& Langer model of BLM tension, suggesting that BLM tension can arise from several sources and can even vary between monolayer leaflets in apparently `tensionless' BLMs \cite{schmidEPL2011}.  While this has led to new BLM models based on renormalized elastic theory \cite{branniganBioJ2006} or a viscoelastic tensor-based approach \cite{arroyoPRE2009}, these are not conclusive and the issue is still unresolved.

Here we present a new model of membrane dynamics that includes a physical description of a `thick' membrane comprising two monolayers, with the freedom to vary parameters between the monolayers.  The thickness is explicitly included by evaluating the boundary conditions at the membrane surface, rather than on the membrane midplane, as previously considered.  We describe the dynamics in terms of intuitive modes of motion of the bilayer, including a form of bilayer relaxation not noted in previous models.  For a symmetric bilayer, the model generates dispersion relations that can be studied analytically and are associated with the modes of bilayer motion.

\section{Method}
\subsection{Geometry}
\begin{figure}[ht]
\centering
\includegraphics[width=\linewidth,clip]{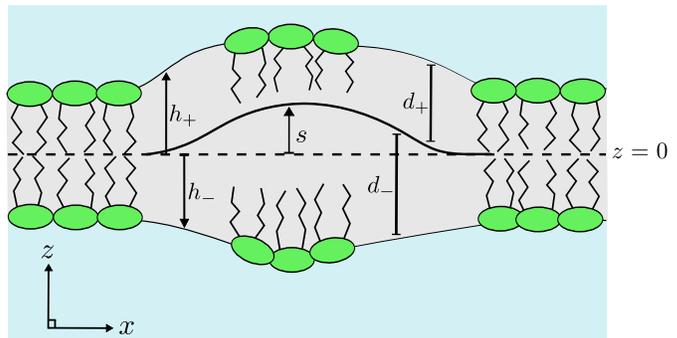}\caption{A 2D projection of the schematic geometry of our bilayer model.\label{fig1}}
\end{figure}

We model an asymmetric BLM by varying the monolayer thickness, membrane viscosity and area compressibility between the monolayer leaflets, which couples the modes of bilayer motion, complicating the dissipative relaxation of an asymmetric BLM.  We consider a planar BLM suspended in viscous fluid, with the unperturbed bilayer normal parallel to the $z$-axis and intermonolayer surface spanning the $x$-$y$ plane at $z=0$, as shown in Fig.~\ref{fig1}.  The upper $\left(+\right)$ and lower $\left(-\right)$ bilayer surfaces are described by the height functions $h_{+}\left(\mathbf{r}\right)$ and $h_{-}\left(\mathbf{r}\right)$:  
\begin{subequations}
\begin{align}
h_{+}\left(\mathbf{r}\right)&=d_{+}\left(\mathbf{r}\right)+s\left(\mathbf{r}\right)\\
h_{-}\left(\mathbf{r}\right)&=-d_{-}\left(\mathbf{r}\right)+s\left(\mathbf{r}\right),
\end{align}
\end{subequations}
where $d_{\pm}\left(\mathbf{r}\right)$ is the thickness (strictly positive) of a monolayer and $s\left(\mathbf{r}\right)$ is the height of the intermonolayer surface.  We define $\mathbf{r}\equiv\left(x,y\right)$.  This explicit description of a `thick' bilayer is a distinctive feature of our model and better reflects the physical nature of a BLM. 
   
\subsection{Free Energy}
The BLM description contains three degrees of freedom; the two monolayer thicknesses and the height of the intermonolayer surface.  By comparison, the Seifert \& Langer model considers two degrees of freedom, the bilayer height and out of phase variation of the monolayer density, where the average density mode is neglected within the non-inertial framework \cite{SeifertEPL1993}.  The free energy of each monolayer $F_{\pm}$ is modelled after the Helfrich-Canham Hamiltonian in the Monge representation \cite{Boalbook2002} with, in principle, different elastic constants for each monolayer:
\begin{subequations}
\begin{align}\label{F.E}
F&=F_{+}+F_{-}+F_{frame};\qquad\text{where,}\\
F_{\pm}&=\tfrac{1}{2}\int\left[\kappa_{b\pm}\left(\nabla^{2}h_{\pm}\right)^{2}+\gamma_{s\pm}\left(\nabla h_{\pm}\right)^{2}\right.\nonumber\\
&\left.+\kappa_{A\pm}\left(\frac{d_{\pm}-d_{0\pm}}{d_{0\pm}}-d_{0\pm}\,\nabla^{2}s\right)^{2}\right]dA,\\
F_{frame}&=\gamma_{fr}\left[\nabla\left(h_{+}+h_{-}\right)\right]^{2}/2,
\end{align}
\end{subequations}
where $\kappa_{b}$ is the membrane bending rigidity, $d_{0\pm}$ is the unperturbed monolayer thickness and $\kappa_{A}$ is the area compressibility.  The second term in the area compressibility couples the curvature of the membrane and the thickness of the monolayers \cite{SeifertAiP1997}.  The surface tension $\gamma_{s}$ restricts variations in the monolayer/water interfacial area while we introduce an additional frame tension term to restrict changes in the total membrane area.  The contribution to the free energy from the frame tension $\gamma_{fr}$ cannot be split between leaflets and is a property of the whole bilayer.  This description of the BLM covers every form of deformation of the bilayer.

\subsection{Dynamics}
The surrounding fluid is assumed to be incompressible ($\nabla\cdot\mathbf{v}=0$) and in the non-inertial Stokes limit (as used by Seifert \& Langer \cite{SeifertEPL1993}), where $\mathbf{v}$ is the fluid velocity \cite{hombergEPL2012}.  This simplifies the corresponding Navier-Stokes equation to 
\begin{equation} 
\nabla p=\eta\,\nabla^{2}\mathbf{v}.\label{NSeqnS}
\end{equation}           
where $p$ is the fluid pressure and $\eta$ is the viscosity.
Each monolayer obeys the continuity equation,
\begin{equation}\label{conteqn}
\frac{\partial n}{\partial t}+\boldsymbol{\nabla}_{\bot}\cdot\left(n\,\mathbf{v}_{l}\right)=0
\end{equation}   
where $\mathbf{v}_{l}$ is the (two-dimensional) velocity of the lipids and $n$ is the two dimensional number density of lipids within the monolayer.  The corresponding area per lipid is $a=1/n$.  We can assume that the volume $v_{0}=a\,d$ per lipid is fixed as this will relax much faster than perturbations to the area or thickness \cite{branniganBioJ2006}.  For small perturbations in $a,d$, and $\mathbf{v}_{l}$, the continuity equation, linking monolayer thickness changes with the flow of lipids within the leaflet becomes;
\begin{equation}\label{cont1:dyn}
\partial_{t}\delta d_{\pm}=-d_{0\pm}\boldsymbol{\nabla}_{\bot}\cdot\mathbf{v}^{\pm}_{l}
\end{equation}
 to lowest order in all perturbations, where $d_0$ is the unperturbed monolayer thickness and $\delta d$ is the perturbation to that thickness. 
\subsection{Boundary Conditions}
The surrounding fluid exerts a shear stress on the monolayer-fluid interface, denoted $T^{\pm}_{iz}$, where $i=x,y$.  Friction also arises at the intermonolayer surface due to the relative velocities of the two monolayers.  We treat the BLM as coupled two-dimensional fluid monolayers, each of which obeys a Navier-Stokes equation.  Including body forces and assuming Stokes flow, the Navier-Stokes equation for each monolayer is 
\begin{equation}\label{NSeqnL}
-d_{0\pm}\boldsymbol{\nabla_{i}}\frac{\delta F}{\delta d_{\pm}}+\mu_{\pm}{\nabla}^{2}\mathbf{v}^{\pm}_{i\,l}\pm \mathbf{T}^{\pm}_{iz}\mp b\left(\mathbf{v}^{+}_{i\,l}-\mathbf{v}^{-}_{i\,l}\right)=0,
\end{equation}       
where $b$ is the coefficient of intermonolayer friction and $\mu_{\pm}$ is the viscosity of the monolayer.  The first term represents the pressure within each monolayer.
In the Stokes approximation the forces across each interface (monolayer/fluid and monolayer/monolayer) balance, which leads to the normal force balance;
\begin{equation}\label{vfbl}
\frac{\delta F}{\delta h_{+}}+\frac{\delta F}{\delta h_{-}}=\mathbf{T}^{+}_{zz}+\mathbf{T}^{-}_{zz}.
\end{equation}
We prescribe non-slip and impermeable boundary conditions at each monolayer-fluid interface at $z=\pm h_{0}$.  The boundary conditions are evaluated at the membrane surface, in contrast to previous studies, where they are evaluated at membrane midplane \cite{SeifertEPL1993,hombergEPL2012}.  This more physical description will capture the membrane thickness:     
\begin{subequations}\label{eq1}
\begin{align}
v^{\pm}_{z}\left(\mathbf{r},z=\pm h_{0}\right)&=\pm\partial_{t}h_{\pm}\left(\mathbf{r}\right)\\
\mathbf{v}^{\pm}\left(\mathbf{r},z=\pm h_{0}\right)&=\mathbf{v}^{\pm}_{l}\left(\mathbf{r}\right).
\end{align}
\end{subequations}
The final boundary condition ensures continuity of the normal component of the velocity across the monolayer-monolayer interface
\begin{equation}\label{eq2}
\partial_{t}h_{+}-\partial_{t}d_{+}=\partial_{t}h_{-}-\partial_{t}d_{-}.
\end{equation}
The dynamics are fully specified by Navier-Stokes equations and the boundary conditions and we require that the velocity and any deviation in the pressure in the surrounding fluid vanish far from the bilayer $\left(\mathbf{v}^{\pm}\rightarrow0 \text{ as } z \rightarrow\pm\infty\right)$.
\subsection{Calculation of Dynamic Modes}
We next recast the bilayer description into collective degrees of freedom that, together with the intermonolayer surface height $s$, reflect all possible modes of bilayer movement:
\begin{eqnarray}\label{eqone}
\bar{h}&=&h_{+}+h_{-}\\
u&=&h_{+}-h_{-}.\nonumber
\end{eqnarray}
Here $u$ describes peristaltic undulations in the bilayer and $\bar{h}$ describes whole bilayer undulations.  If $\bar{h}$ and $s$ undulate with the same wavelength and amplitude, then the bilayer undergoes a thickness-preserving undulation, comparable to the height undulations of previous dynamic models \cite{BrochardJDP1975,SeifertEPL1993}.  This recasts the free energy framework in terms of $\bar{h}$, $u$ and $s$, rather than $d_{\pm}$/$h_{\pm}$ and $s$.  

We expand in a Fourier series,
\begin{equation}
\bar{h}=\sum_{q}\bar{h}_{q}\left(t\right)e^{-i\mathbf{q}\cdot\mathbf{r}},
\end{equation}
and similarly for $u\left(\mathbf{r}\right)$ and $s\left(\mathbf{r}\right)$.  Here $\mathbf{q}\equiv(q_{x},q_{y})$ is the Fourier wavelength of the undulation.
The dynamical variables then obey
\begin{equation}\label{mat}
\partial_{t}\begin{pmatrix}\bar{h}_{q}\\u_{q}\\s_{q}\end{pmatrix}=-M_{q}\begin{pmatrix}\bar{h}_{q}\\u_{q}\\s_{q}\end{pmatrix}\\
\end{equation}
where
\begin{equation}
M_{q}=\left[\begin{array}{ccc}M_{11}&0&0\\M_{21}&M_{22}&M_{23}\\M_{31}&M_{32}&M_{33}\end{array}\right].
\end{equation}
The matrix $M_{q}$ contains all the fundamental information of the system (to linear order).  The solutions are given by 
\begin{equation}
\mathbf{\Lambda}_{i}\left(t\right)=\Lambda_{0}\mathbf{\hat{e}}_{i}\,e^{-\omega_{i}\left(q\right)t}\quad\textnormal{where}\,\,\,i=1,2,3.
\end{equation}
where $\mathbf{\Lambda}_{0}$ is the initial amplitude of the bilayer deformation represented by the $i$th eigenmode, which has eigenvalue (decay rate or dispersion relation) $\omega_{i}\left(q\right)$, and eigenvector $\mathbf{\hat{e}}_{i}$.  

\begin{table}[ht]
\centering{\caption{Table of parameters.\label{tab:1}}
\begin{tabular}{ccc}\\\hline\hline
Symbol & Value & Reference \\ [0.5ex]  \hline
$\kappa_{b}$ & $40\,k_{B}T$ & \cite{Boalbook2002}\\
$\gamma_{s}$ & $0.1\,\text{mN}\text{m}^{-1}$ & \cite{SeifertAiP1997}\\
$\gamma_{fr}$ & $0.3\,\text{mN}\text{m}^{-1}$ & \cite{SeifertAiP1997}\\
$\kappa_{A}$ & $0.14\,\text{N}\text{m}^{-1}$ & \cite{SeifertAiP1997}\\
$d_{0}$ & $2\,$nm & \cite{Boalbook2002}\\
$\mu$ & $10\,$nPa$\,$ms & \cite{SeifertEPL1993}\\
$\eta$ & $1\,$mPa$\,$s & \cite{SeifertEPL1993}\\
$b$ & $ 10\,\text{MPa}\,\text{s}\,\text{m}^{-1}$ & \cite{OtterBioJ2007}\\
[1ex]\hline
\end{tabular}}
\end{table}

\section{Results}
Since the matrix $M_{q}$ is symmetric in $q_{x},q_{y}$ we assume $q_{y}=0$ without loss of generality.  We first consider a symmetric bilayer, where the BLM parameters are identical in each monolayer leaflet.  This simplifies $M_{q}$ into diagonal components and one off-diagonal component ($M_{31}$) which couples the whole bilayer mode, $\bar{h}$ and intermonolayer surface height $s$.  The diagonal components provide analytic expressions for the eigenvalues $\omega_{i}\left(q\right)$.  
\subsection{Peristaltic mode}
The first mode has eigenvalue $\mathbf{\hat{e}}={u_{q}}$ and is a pure peristaltic dynamic mode describing undulations in the bilayer where the thicknesses of the monolayers undulate in phase.  The decay rate is 
\begin{equation}
\omega_{u}\left(q_{x}\right)=M_{22}=\frac{q_{x}\,\kappa_{A}}{\mu q_{x}+2\,\eta}=\frac{q_{x}\,\kappa_{A}}{\eta\left(\ell_{SD}q_{x}+2\right)},
\end{equation}
in which the area compressibility is damped by the viscosities of the solvent and lipids.  The effect of the membrane viscosity can be understood when the eigenvalue is expressed in terms of the Saffman-Delbr\"{u}ck length $\ell_{SD}=\mu/\eta$ \cite{SaffmanPNAS1975}.  The Saffman-Delbr\"{u}ck length represents the length scale below which the membrane viscosity dominates the surrounding fluid viscosity.  For wavelengths larger than $\ell_{SD}$ $\left(q\,\ell_{SD}\ll1\right)$ the fluid drag dominates while for shorter wavelengths the in-plane monolayer viscosity dominates and the damping is independent of wavelength.  For the membrane viscosity ($\mu\approx10\,\text{nPa}\,\text{m}\,\text{s}$) and the fluid viscosity ($\eta\approx1\,\text{mPa}\,\text{s}$) that are used here $\ell_{SD}\approx\,0.1\mu\text{m}$ therefore $\omega_{u}\left(q\right)\simeq\kappa_{A}/\mu$ for smaller scale undulations.  The Saffman-Delbr\"{u}ck length emerges from considering the surrounding fluid moved by thickness fluctuations, an effect neglected previously \cite{SeifertEPL1993,OtterBioJ2007}.  The limiting value of this dispersion relation (at high wavenumber $\omega_{u}\left(q\right)\sim10^{7}$s$^{-1}$) is comparable to the experimental relaxation rate ($100$ns) for thickness fluctuations measured by Woodka \emph{et al.} \cite{woodkaPRL2012}. 

This dispersion relation is comparable to the density difference region of an eigenvalue found by Seifert and Langer \cite{SeifertEPL1993,SeifertBC1994}.  Seifert and Langer found a mixed eigenvalue where terms representing the peristaltic bilayer dynamic mode dominate for a particular region of $q$ values given in their paper as $\gamma_{2}$ for $q\gg2\,\eta\kappa_{A}/(b\,(\kappa_{b}+2\,d_{0}\kappa_{A}))$, as the combined density and height variables used are coupled when describing this form of bilayer motion.  We find a single eigenvalue due to the decomposition of membrane motion into the collective degrees of freedom (Eq.~\ref{eqone}).    

\subsection{Ripple mode}
We next consider a mixed dynamic mode that couples the bilayer mode $\bar{h}$ and the internal surface mode $s$, although the contribution from the latter has a very weak wavelength dependence and is slightly suppressed at low $q$.  The eigenvector is given by 
\begin{align}
\mathbf{e}_{\bar{h}s}&=\left[\begin{array}{c}1\\0\\\frac{M_{31}}{M_{11}-M_{33}}\end{array}\right]\\
&\text{where}\nonumber\\
M_{31}&=\frac{M_{33}}{2\left(d_{0}^{2}\,q_{x}^{2}-1\right)}-\frac{M_{11}}{2}.\nonumber
\end{align}
In this dynamic mode, the internal surface $s$ undulates in phase with the bilayer surface in order to preserve the thickness of each monolayer, and the decay rate is 
\begin{equation}
\omega_{\bar{h}s}\left(q_{x}\right)=M_{11}=\frac{(\gamma_{s}+2\gamma_{fr}+\kappa_{b}q_{x}^{2})q_{x}e^{-q_{x}\,d_{0}}}{2\eta}.
\end{equation}
Relaxation is driven by the membrane bending rigidity ($\kappa_{b}$) and the surface and frame tensions, and resisted by the fluid viscosity.  This relation is equivalent to the ripple-type dynamic mode of Brochard \& Lennon \cite{BrochardJDP1975}.  Here the dynamic mode is embodied in a single eigenvalue, whereas in previous models the dynamic mode dominates within a larger, more complex eigenvalue (the 'fast' mode in Seifert \& Langer) in the shorter wavelength range $q\approx10^{-2}-1\text{nm}^{-1}$.  

\subsection{Internal mode}
\begin{figure}[ht!]
	\centering{\includegraphics[width=\linewidth,clip]{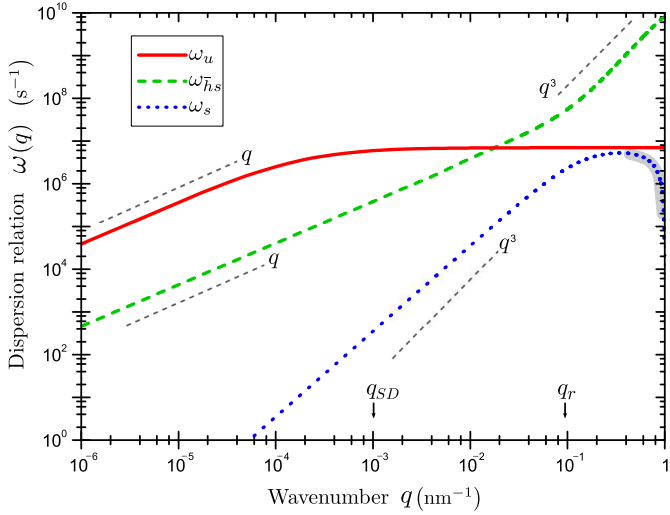}}
	\caption{Dispersion relations ($\omega\left(q\right)$) as a function of the Fourier wavenumber ($q$).  The wavenumbers $q_{SD}=2\pi/\ell_{SD}$ and $q_{r}=2\pi/\ell_{r}$ associated with the Saffman-Delbr\"{u}ck length $\ell_{SD}$ and the new length scale $\ell_{r}$ are shown.  Power laws are shown for comparison.  The shaded region of $\omega_{s}$ indicates where the Stokes approximation breaks down, leading to the false impression of a maximum value.  The parameters used to generate the figure are shown in Table \ref{tab:1}.\label{fig:dr1}}
\end{figure}

Finally, we find a dispersion relation associated with movement of the internal surface $s_{q}$;
\begin{equation}
\omega_{s}\left(q_{x}\right)=M_{33}=\frac{\kappa_{A}\,q_{x}^{2}\,\left(q_{x}^{2}\,d^{2}_{0}-1\right)}{\eta\left(2\,q_{x}+q^{2}_{x}\ell_{SD}+2\,\ell_{SD}/\ell_{r}^{2}\right)},
\end{equation}
where the length scale $\ell_{r}=\sqrt{\mu/b}$, compares the intermonolayer friction and the membrane viscosity.  This dispersion relation is a balance between the area compressibility ($\kappa_{A}$) and the viscosities $\eta$ and $\mu$ and the intermonolayer friction, $b$.  A similar dependence is seen in the `slipping' region ($q\approx10^{-2}-1\text{nm}^{-1}$) of the slower mode of Seifert \& Langer (denoted by $\gamma_{2}$ in their work).  The length scale $\ell_{r}$ represents the size at which the forces from the membrane viscosity and the intermonolayer friction on a monolayer inclusion balance.  For a typical bilayer ($\mu\approx10\,\text{nPa}\,\text{m}\,\text{s}$, $b\approx10\,\text{MPa}\,\text{s}\,\text{m}^{-1}$ ) $\ell_{r}\approx10\text{nm}\ll \ell_{SD}$ as the intermonolayer friction is much larger than the membrane viscosity, although current estimates of $b$ do show large variation \cite{OtterBioJ2007}.  The movement of lipids within the monolayer also drags the surrounding solvent.  The dispersion relation tends to zero as $qd_{0}\rightarrow1$, where the Fourier wavelength approaches the membrane thickness.  This unphysical behaviour is expected in this limit, as the Stokes approximation begins to break down.  At short wavelengths, the inclusion of inertial effects have been found to lead to a $q^{4/3}$ dependence \cite{langevincis1981}.  The Stokes approximation breaks down for only the internal mode $\omega_{s}$ as this contains the least dissipation of motion in the surrounding fluid of all modes and hence the inertia/dissipation balance that leads to the Stokes approximation is disrupted before it will be in the remaining modes.  This relevant portion of the figure is marked on figures \ref{fig:dr1} \& \ref{fig:dr2}.          

The dispersion relations are shown in Fig.~\ref{fig:dr1}.  The dispersion relation $\omega_{s}$ associated with the internal monolayer/monolayer surface has the smallest magnitude for the entire range of $q$, meaning that this type of bilayer motion will be the slowest to relax.  At long wavelengths the fastest mode is the peristaltic mode $\omega_{u}$,  before crossing over to the thickness preserving ripple mode $\omega_{\bar{h}s}$.  This crossover occurs at $q=10^{-2}\,\text{nm}^{-1}$ equivalent to an undulation wavelength of $0.1\,\mu\text{m}$, which is comparable with the crossover value between the `slipping' and `ripple' dynamic modes found in previous models \cite{SeifertEPL1993,SeifertBC1994}, simulations \cite{GoetzPRL1999} and experiments \cite{shkulipaPRL2006}.   

The eigenvalues $\omega_{\bar{h}s}$ and $\omega_{u}$ exchange their relative magnitudes as the Fourier wavelength $q$ is increased, but the eigenvectors show an extremely weak dependence on $q$ and so that the nature of each mode does not change as $q$ increases.  In the model first presented by Seifert \& Langer \cite{SeifertEPL1993,SeifertBC1994}, two dynamic modes interchanged their character depending on $q$ range, which makes it difficult to interpret the bilayer motion.  
\subsection{Asymmetric BLM}
\begin{figure}[ht!]
	\centering{\includegraphics[width=\linewidth,clip]{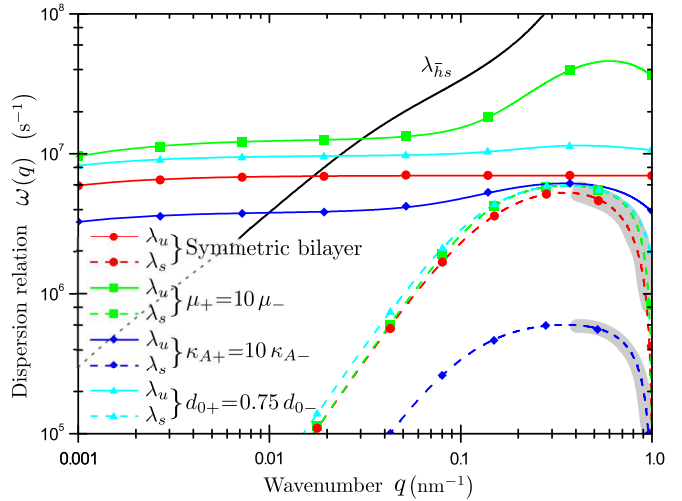}}
	\caption{A log-log plot of the dispersion relations ($\omega\left(q\right)$), showing the effect of increasing asymmetry.  The dispersion relation for the ripple mode undulation $\omega_{\bar{h}s}\left(q\right)$ is unchanged by the asymmetry considered in this study.  The parameters used to generate the figure are shown in Table \ref{tab:1}.\label{fig:dr2}}
\end{figure}

The results considered so far are for a symmetric BLM, which is clearly an idealised case, since biological membranes are often asymmetric \cite{Nelsonbook2004}.  Varying the bending modulus, $\kappa_{b}$ and the stretching moduli, $\gamma_{b}$ and $\gamma_{fr}$, is non-trivial.  Here we vary the area compressibility, $\kappa_{A}$, the relative leaflet thickness, $d_{0}$, and the membrane viscosity, $\mu$, between leaflets.  In this case, the matrix $M_{q}$ does not simplify, and expressions for the eigenvalues do not lead themselves to analytical inspection, with the exception of $M_{11}$, which is unchanged.  The exponential in $M_{11}$ is unaffected as the $d_{0}$ comes from a boundary condition which depends on the whole bilayer thickness, while we consider variation in the relative leaflet thicknesses.  The changed terms of the matrix are given by
\begin{align}
M_{21}&=q_{x}\left[\kappa_{A-}\alpha^{2}\left(2\,b+2\eta q_{x}+\mu_{+} q_{x}^{2}\right)\right.\\
&\left.-\kappa_{A+}\left(2\,b+2\eta q_{x}+\mu_{-} q_{x}^{2}\right)\right]/2\,C_{1}\nonumber\\
M_{22}&=-q_{x}\left[\kappa_{A-}\alpha^{2}\left(2\,b+2\eta q_{x}+\mu_{+} q_{x}^{2}\right)\right.\\
&\left.+\kappa_{A+}\left(2\,b+2\eta q_{x}+\mu_{-} q_{x}^{2}\right)\right]/2\,C_{1}\nonumber\\
M_{23}&=q_{x}\left[\kappa_{A-}\alpha^{2}\left(d_{0-}^{2}q_{x}^{2}-1\right)\left(2\,b+2\eta q_{x}+\mu_{+} q_{x}^{2}\right)\right.\\
&\left.-\kappa_{A+}\left(\alpha^{2}\,d_{0-}^{2}q_{x}^{2}-1\right)\left(2\,b+2\eta q_{x}+\mu_{-} q_{x}^{2}\right)\right]/2\,C_{1}\nonumber\\
M_{31}&=q_{x}e^{-q_{x}d_{0}}\left\{q_{x}\,\eta\,e^{q_{x}d_{0}}\left[\kappa_{A-}\alpha^{2}\left(\mu_{+}q_{x}+2\,\eta\right)\right.\right.\\
&\left.\left.+\kappa_{A+}\left(\mu_{-}q_{x}+2\,\eta\right)\right]\right.\nonumber\\
&\left.-\alpha^{2}\left(\gamma_{s}+2\gamma_{fr}+\kappa_{b}q_{x}^{2}\right)C_{1}\right\}/4\,\eta\,C_{1}\nonumber\\
M_{32}&=q_{x}^{2}\left[\alpha^{2}\kappa_{A-}\left(2\,\eta+\mu_{+}q_{x}\right)\right.\\
&\left.-\kappa_{A+}\left(2\,\eta+\mu_{-}q_{x}\right)\right]/4\,C_{1}\nonumber\\
M_{33}&=q_{x}^{2}\left[\alpha^{2}\kappa_{A-}\left(d_{0-}^{2}q_{x}^{2}-1\right)\left(2\,\eta+\mu_{+}q_{x}\right)\right.\\
&\left.+\kappa_{A+}\left(\alpha^{2}\,d_{0-}^{2}q_{x}^{2}-1\right)\left(2\,\eta+\mu_{-}q_{x}\right)\right]/2\,C_{1}\nonumber
\end{align}
where
\begin{align}
\alpha&=d_{0+}/d_{0-}\\
C_{1}&=\alpha^{2}\left[4\,b\,\eta+q_{x}\left(\mu_{-}+\mu_{+}\right)\left(2\,\eta\,q_{x}+b\right)\right.\\
&\left.+4\,\eta^{2}\,q_{x}+\mu_{+}\mu_{-}q_{x}^{3}\right].\nonumber
\end{align}
The numerically calculated eigenvalues are shown in figure \ref{fig:dr2}.  

The position of the crossover wavelength between $\omega_{\bar{h}s}\left(q\right)$ and $\omega_{u}\left(q\right)$ depends upon the degree of asymmetry, as do the dispersion relations $\omega_{u}\left(q\right)$ and $\omega_{s}\left(q\right)$.  Introducing asymmetry has a seemingly small effect on the magnitude of the dispersion relations, but does influence the character of the modes, as seen in the eigenvectors shown in figure \ref{fig:dr2}.  The dispersion relations $\omega_{s}\left(q\right)$ and $\omega_{u}\left(q\right)$ become mixed modes associated with both $u$ and $s$; so that a peristaltic excitation of an asymmetric BLM will relax both by peristaltic undulations ($u$) and by movement of the internal surface ($s$).  If the dynamic modes are coupled, then an asymmetric membrane would relax at the rate of the fastest dynamic mode, as this would provide the fastest return to equilibrium.  This changes the dissipative dynamics of BLM in experimental membranes.  

\section{Conclusions}
We have constructed a new model and framework for the dynamics of BLMs that generalises to asymmetric membranes.  The framework provides an intuitive description of both the BLM and the forms of bilayer motion.  We explicitly include the effect of thickness fluctuations on the surrounding fluid and find a limiting value in the dispersion relation associated with these thickness fluctuations that agrees well with recent experiments \cite{woodkaPRL2012}.  Two length scales emerge from the dispersion relations: the previously well studied Saffman-Delbr\"{u}ck length $\ell_{SD}$, and a new lengthscale, $\ell_{r}$ a monolayer equivalent to the Saffman-Delbr\"{u}ck length which balances internal monolayer viscosity with internal friction.  The additional length scale emerges as the associated form of bilayer motion ($s$) is considered uncoupled from the other forms of bilayer motion.  Asymmetry does not significantly effect the relaxation dynamics of the bilayer but instead couples the dynamic modes, meaning an asymmetric membrane (e.g. most biological membranes) will relax by a combination of dynamic modes.  An asymmetric surface tension or bending rigidity may change the dynamics more significantly but represents a significant extension to the model.  Biological membranes contain numerous embedded molecular complexes. The effect of membrane behaviour on these complexes has been studied elsewhere \cite{HuangBioJ1986}, and their inherent asymmetry (either by their conformation or by their interactions with the surrounding lipids) will affect membrane dynamics. Complex embedded structures such as a transmembrane ion channel could be modelled within our existing membrane framework, but would be best treated within a fully three dimensional description, which would make obtaining analytic solutions difficult.  The perfect symmetry required to observe ‘pure’ hydrodynamic modes presents a challenge for experimentalists. However, the change in the wavelength of the crossover between the slowest relaxing modes could provide an opportunity for experimental verification.  Recent advances in experimental methods allow for the preparation of intentionally asymmetric membranes \cite{ElaniCC2015}.  Observations of the flucuation spectra for these membranes may validate the predictions of the model.

\acknowledgments
This work was supported by EPSRC grant no. 548622/1 as part of the White Rose Training Centre in Physical Methods and Life Sciences.

\end{document}